\documentstyle[preprint,aps,epsf]{revtex}
\tightenlines
\def\Journal#1#2#3#4{{#1} {\bf #2}, #3 (#4)}

\def\AP{{\em Ann. Phys.}}

\def\IJMPA{{\em Int. J. Mod. Phys.} A}
\def\JFA{{\em J. Funct. Anal.}}
\def\JMP{{\em J. Math. Phys.}}
\def\JPA{{\em J. Phys.} A}

\def\NP{{\em Nucl. Phys.}}
\def\NPB{{\em Nucl. Phys.} B}

\def\PRA{{\em Phys. Rev.} A}

\def\PRD{{\em Phys. Rev.} D}
\def\PRL{\em Phys. Rev. Lett.}
\def\PTP{{\em Prog. Th. Phys.}}

\newcommand{\be}{\begin{equation}}
\newcommand{\ee}{\end{equation}}
\newcommand{\bea}{\begin{eqnarray}}
\newcommand{\eea}{\end{eqnarray}}

\newcommand{\hf} {{1\over2}}

\newcommand{\nonu}{\nonumber\\}
\def\ra{\rangle}
\def\la{\langle}

\def\ca{{\cal A}}
\def\dt{\Delta\tau}

\def\eq#1{(\ref{#1})}

\begin{document}
\title{Path Integral for the Dirac Equation}
\author{Janos Polonyi}
\address{Laboratory of Theoretical Physics, Louis Pasteur University\\
3 rue de l'Universit\'e 67087 Strasbourg, Cedex, France\\
and\\
Department of Atomic Physics, L. E\"otv\"os University\\
P\'azm\'any P. s\'et\'any 1/A 1117 Budapest, Hungary}
\date{\today}
\maketitle
\begin{abstract}
A c-number path integral representation is constructed for the 
solution of the Dirac equation. The integration is over the 
real trajectories in the continuous three-space and other 
two canonical pairs of compact variables controlling the 
spin and the chirality flips. 
\end{abstract}

The path integral representation of the quantum amplitudes
\cite{feyn} has become an essential device in the development of
an intuitive picture about quantum physics, the
organisation of the perturbation expansion, and 
in establishing non-perturbative approximations. A serious
limitation of the functional methods arises from
the difficulties in extending the path integral formalism
for relativistic fermions. There have been numerous proposals
to solve this problem.
The original attempt at a path integral representation
for the solution of the 1+1 dimensional massless 
Dirac equation was a random walk along the line cone
of a space-time lattice with a Poisson distributed helicity 
flips \cite{feyn,latt}. The method was soon generalised for 
3+1 dimensions \cite{thpo} and the highly complicated 
time ordered products of the Dirac matrices was dealt with by 
the integration over Grassman-valued trajectories \cite{gasc}. 
The propagation along the space-time trajectories and in the
spinor space was separated and the latter was treated
by brute force in refs. \cite{effa}. The propagation can
be described in a five dimensional manifold and the resulting
path integrals resemble the proper time formalism \cite{prop}.
Guided by supersymmetry \cite{susy} the Grassman variables 
were introduced combined with the proper time \cite{gras}.
We aim at a path integral formalism which
stays as close as possible to the physics of the problem
(does not involve proper time, the continuous trajectories are 
parametrized by the real time), and is based on c-number integration
to allow the use of non-perturbative methods, such as the semiclassical
approximation for the fermion propagator. Such a formalism,
employed with the second quantization might bring bosons and
fermions on equal footing. 

The goal of the present paper is the path integral
solution of the first quantized Dirac equation
\be
i\hbar\partial_t\Psi=H_D\Psi=\left[\left(-ic\hbar\partial+eA\right)
\cdot\alpha+mc^2\beta-eV\right]\Psi,
\label{dham}
\ee
where $\Psi(x,t)$ denotes the Dirac bi-spinor, $(A(x,t),V(x,t))$
is an external gauge field and $\alpha_i=\gamma_0\gamma_j$,
$\beta=\gamma_0$ are Dirac matrices. 
Before embarking the technical problems to resolve in finding
the path integral one is faced with the usual 
inconsistency of the first quantized quantum mechanics, 
namely the existence of states with negative energy. This appears
in our context as the ill defined nature of the time 
evolution operator $e^{-itH_D/\hbar}$,
even for imaginary time. We avoid this problem by
the replacement $H_D\to H_{cutoff}=H_D+p^2/2M$ in the exponential. 
The mass parameter $M$ introduced in this manner
plays the role of an UV cutoff because 
this modification is acceptable only for $p<Mc$. The cutoff
is implemented in the path integral formalism by the small
but finite time step $\Delta t$. We shall find the relation
$Mc^2=\kappa\hbar/\Delta t$ where $\kappa$ is a real positive
constant, indicating that the time evolution
operator is well defined when $Im(t\Delta t)<0$, for 
the time axis which is tilted as $t\to te^{i\alpha}$,
$-\pi/2<\alpha<0$. The choice $\alpha=-\pi/4$
will serve as the safe starting point to establish
the convergence of the path integral expressions
introduced below by means of the analytic continuation
either to real or imaginary time.

The problems to solve in the path integral formalism
is (i) to generate the first order derivatives in the space
coordinates and (ii) to treat the spinor indices. 
The problem (i) is circumvented within the framework of the Wiener-type 
path integration by using the bare cutoff hamiltonian, $H_{cutoff}$. 
The main difference with the non-relativistic case
is the unusual cutoff dependence \cite{gopo}. (ii) is solved 
by a well known method, the introduction of
additional continuous degrees of freedom for the representation of
the dynamics in the Dirac space \cite{spin}.

In the first step we construct the integration over the three-space
trajectories by keeping the spinor dynamics in its algebraic form.
For this end we introduce the infinitesimal time propagator 
$G(x,y;\dt)=e^{i\ca(x,y;\dt)}$ defined as
\be
\Psi(x,\tau+\dt)=\int dyG(x,y;\dt)\Psi(y,\tau),\label{infev}
\ee
where $\tau=tc$. One may motivate the choice of $\ca(x,y;\dt)$
for free massless fermions in the following manner. The self 
reproducibility,
\be
e^{i\ca(x,y;\tau_1+\tau_2)}=\int dz
e^{i\ca(x,z;\tau_1)}e^{i\ca(z,y;\tau_2)},\label{gblock}
\ee
can easiest be satisfied by using quadratic expression in the 
coordinates for $\ca(x,y;\dt)$.
The translation and rotation symmetries require the functional form
\be
\ca(x,y;\dt)={A(\dt)\over2}(x-y)^2+B(\dt)(x-y)\cdot\tilde S+C,
\ee
$\tilde S_j$ standing for a vector which transforms as the spin
under rotations. Since the equation of motion contains
no $\hbar$ we keep $\ca$ $\hbar$-independent. Then
the choice $A=a/\dt^2$ and $B=b/\dt$ follows by dimensional 
argument. We can carry out the integral in \eq{infev} after expanding
the amplitude in the right hand side at $y=x$ one arrives at 
\be\label{neqd}
\Psi(x,\tau+\dt)=\int dy
e^{i\left[{a\over2\dt^2}(x-y)^2-{b\over\dt}(x-y)\cdot\tilde S+C\right]}\Psi(y,t)\\
=\biggl[1-{\dt b\over a}\tilde S\cdot\partial\biggr]\Psi(x,t)
\ee
for the appropriate choice of the constant $C$ and having dropped 
the contributions $O(\dt^2)$.
This agrees with the massless Dirac equation for $S_j=\alpha_j$, $a=b=\kappa$
if the characteristic length of the amplitude is long
enough, $\dt<<\hbar/|p|$, to ignore the terms $O(\dt^2)$. The mass represents
a mixing between the chiral spinors and can be taken into account
by the modification $G\to Ge^{-i\beta\dt/\lambda}$, $\lambda=\hbar/mc$ and 
$\dt<<\lambda$. The resulting path integral in the absence of the external field is
\be
G(x,y;\tau)=\prod_j\left\{\int dx_je^{
i\kappa\left[{1\over2}\left({x_j-x_{j+1}\over\dt}\right)^2
-{x_j-x_{j+1}\over\dt}\cdot\alpha\right]+iC(\dt)}
e^{-i\dt{mc\over\hbar}\beta}\right\}
\ee
for a suitable chosen $C(\dt)$. One introduces the formal notation 
$\lim\limits_{\dt\to0}\prod_j\int dx_je^{iC(\dt)}=\int D[x]$ and can writes
\be
G(x,y;\tau)=\int D[x]T\left\{e^{i{\kappa\over\dt}\int d\tau
\left[\hf\left({dx\over d\tau}\right)^2-{dx\over d\tau}\cdot\alpha\right]}
e^{-i{mc\over\hbar}\int d\tau\beta}\right\}.\label{pimatr}
\ee
The convergence of the path integral can be established 
by starting with the time axis tilted by $-\pi/4$ as explained
in the introduction and performing the analytical continuation 
for real time. The path integral can easily be computed for the 
massless case,
\be
G(x,y;\tau)=\int{d^3p\over(2\pi)^3}e^{{i\over\hbar}p\cdot(x-y)}
e^{-i\tau\left[{\dt\over2\kappa}\left({p\over\hbar}\right)^2
-{p\over\hbar}\cdot\alpha\right]+c},
\ee
in agreement with the form of $H_{cutoff}$.

The typical trajectory in \eq{pimatr} is more regular than
in the non-relativistic path integral because the linearly 
diverging coupling constant $\dt^{-1}$ of the action
suppresses the fluctuations. The divergence in the exponent of the 
integrand is not a real surprise, one is accustomed to find diverging 
parameters in the bare, cutoff quantum field theories and
even in non-relativistic quantum mechanics \cite{pol}.
In fact, the cutoff of a renormalizable model is removed in the following
manner: We are given first a regulated path integral involving the lagrangian
$L(\phi,\partial_\mu\phi;g_n)$ and a cutoff $\Lambda$. The
cutoff is then sent to the infinity. The parameters of the lagrangian
become functions of the cutoff, $g_n\to g_n(\Lambda)$ in such a manner
that the regulated path integral with the lagrangian 
$L(\phi,\partial_\mu\phi;g_n(\Lambda))$ yields convergent observables
as $\Lambda\to\infty$.
The fine tuning of the bare parameters is introduced in our case
for the same goal, to
arrive at a cutoff independent, convergent result, \eq{dham}.
One can verify the scaling relations
\be
\la\left({\Delta x\over\dt}\right)^2\ra=\cases{1&$\dt<<\lambda$,\cr
\lambda/\dt&$\dt>>\lambda$}
\ee
confirming the relativistic crossover at $\dt\approx\lambda$ \cite{gopo}.
These scaling laws indicate that the non-differentiable fractal
trajectories of the non-relativistic quantum mechanics are
"regulated" by embedding them into a smoother relativistic propagation.

In the second step we eliminate the time ordered product
of matrices in the spinor space \cite{spin}. For this end
we write the matrices of the Dirac hamiltonian 
in the chiral representation as the direct products 
$\alpha_j=\sigma_j\otimes\tau_3$ and $\beta=\tau_1$, where
$\sigma$ and $\tau$ are the Pauli matrices acting in the
spin and the chirality spaces, respectively. In the matrix formalism
\eq{pimatr} one uses the basis vectors $|m-\hf\ra$ $m=0,1$ in the 
spin and the chirality space. Instead of these vectors the 
dynamics in the four dimensional Dirac space will be followed
by means of the overcomplete and continuous basises
$|\phi\ra$ and $|\chi\ra$, introduced for the spin and the
chirality, respectively,
\be
|\alpha\ra={1\over\sqrt{2\pi}}\sum_{m=0,1} e^{-i\alpha m}|m-\hf\ra,
\ee
with $-\pi\le\alpha<\pi$. 
The bi-spinor $\Psi(x,t)$ is converted into a single component amplitude
by projecting it onto the basis vector $|\phi,\chi\ra=|\phi\ra\otimes|\chi\ra$,
$\Psi(\phi,\chi,x,t)=\la\phi,\chi|\Psi(x,t)\ra$. The time evolution is given by
\bea
\Psi(\phi,\chi,x,\tau+\dt)=\int dyd\phi'd\chi'
G(x,\phi,\chi,y,\phi',\chi';\dt)\Psi(\phi',\chi',y,\tau),
\eea
where 
\be
G(x,\phi,\chi,y,\phi',\chi';\dt)=\la\phi,\chi|e^{-{i\over\hbar}\dt H_{cutoff}}
|\phi',\chi'\ra.
\ee
By the help of the relation
\be
\int\limits_{-\pi}^\pi d\phi\la m'-\hf|\phi\ra\la\phi|m-\hf\ra
=\delta_{m',m},
\ee
$m,m'=0,1$ we find the resolution of the identity 
\be
\int\limits_{-\pi}^\pi d\alpha|\alpha\ra\la\alpha|=1.
\ee
The repeated insertion of this form of the identity into the
time evolution operator yields the path integral where
the trajectories are $x(\tau)$, $\phi(\tau)$ and $\chi(\tau)$.
The Poisson resummation formula
\be
\sum\limits_{\ell=\ell_1}^{\ell_2}f(\ell)=\sum\limits_{n=-\infty}^\infty
\int\limits_{\ell_1-\epsilon_-}^{\ell_2+\epsilon_+}dpf(p)e^{i2\pi np},
\ee
where $\ell,~\ell_j,~n$ are integers and $0<\epsilon_\pm<1$ can be used to 
write the overlap between the basis elements as
\be
\la\alpha|\beta\ra=\sum\limits_{n=-\infty}^\infty
\int\limits_{-\epsilon_-}^{1+\epsilon_+}{dp\over2\pi}
e^{ip(\alpha-\beta+2\pi n)},
\ee
suggesting the interpretation of the variables $\alpha$ and $p$ as
canonical pairs.

We need in \eq{neqd} the matrix elements of the Pauli matrices,
\bea
\la\alpha|\sigma_3|\beta\ra&=&{e^{i(\alpha-\beta)}-1\over2\pi}\nonu
&=&{1\over i\pi}e^{{i\over2}(\alpha-\beta)}\partial_\alpha
e^{-{i\over2}(\alpha-\beta)}\sum_{m=0,1} e^{im(\alpha-\beta)}\nonu
&=&2e^{{i\over2}(\alpha-\beta)}\partial_\alpha
e^{-{i\over2}(\alpha-\beta)}\sum_{n=-\infty}^\infty\int\limits_{-0}^{1+0}
{dp\over2\pi}pe^{ip(\alpha-\beta+2\pi n)}\nonu
&=&2\sum_{n=-\infty}^\infty\int\limits_{-0}^{1+0}
{dp\over2\pi}\left(p-\hf\right)e^{ip(\alpha-\beta+2\pi n)},\nonu
&=&2\sum_{n=-\infty}^\infty\int\limits_{-1+0}^{1-0}
{dp\over2\pi}pe^{i(p+\hf)(\alpha-\beta+2\pi n)},\label{zmip}
\eea
where the Poisson resummation formula was used in the third equation
to write the sum over $m$ as an integral.
The matrix elements of $\sigma_\pm=\sigma_1\pm i\sigma_2$ are given by
$\la\alpha|\sigma_+|\beta\ra=e^{i\alpha}/\pi$,
$\la\alpha|\sigma_-|\beta\ra=e^{-i\beta}/\pi$,
\be
\la\alpha|\sigma_\pm|\beta\ra={1\over\pi}\sum_n\delta_{n,0}
e^{i(n+\hf)(\alpha-\beta)\pm{i\over2}(\alpha+\beta)}f(n),
\ee
where $f(n)$ is an arbitrary function satisfying $f(0)=1$. The Poisson
resummation formula yields
\be
\la\alpha|\sigma_\pm|\beta\ra=2\sum\limits_{n=-\infty}^\infty
\int\limits_{-1+0}^{1-0}{dp\over2\pi}\sqrt{1-p^2}
e^{i(p+\hf)(\alpha-\beta+2\pi n)\pm{i\over2}(\alpha+\beta+2\pi n)},
\label{plmip}
\ee
for the choice $f(n)=\sqrt{1-n^2}$. The application of \eq{zmip} and \eq{plmip}
gives the matrix elements of
$v\cdot\sigma=v_3\sigma_3+(v_+\sigma_-+v_-\sigma_+)/2$, $v_\pm=v_1\pm iv_2$,
\bea
\la\alpha|v\cdot\sigma|\beta\ra&=&2\sum\limits_{n=-\infty}^\infty
\int\limits_{-1+0}^{1-0}{dp\over2\pi}e^{i(p+\hf)(\alpha-\beta+2\pi n)}\nonu
&&\times\left\{v_3p+\sqrt{1-p^2}\left[v_1\cos\left({\alpha+\beta+2\pi n\over2}\right)+
v_2\sin\left({\alpha+\beta+2\pi n\over2}\right)\right]\right\}.
\eea

We are now in the position to rederive \eq{pimatr} in the continuous basis,
\be
G(x_f,\phi_f,\chi_f,x_i,\phi_i,\chi_i;\tau)=
\prod\limits_{k=1}^{N-1}\left(\int dx_kd\phi_kd\chi_k\right)
\prod\limits_{j=1}^N\left(\int{ds_{3,j}\over2\pi}{dt_{3,j}\over2\pi}
\sum\limits_{n_j,N_j=-\infty}^\infty\right)e^{i\ca'},
\ee
as $N\to\infty$, with
\bea
\ca'&=&\sum\limits_{j=1}^N\biggl[\left(s_{3,j}+\hf\right)(\phi_j-\phi_{j-1}+2\pi n_j)
+\left(t_{3,j}+\hf\right)(\chi_j-\chi_{j-1}+2\pi N_j)\nonu
&&+{\kappa\over2}\left({x_j-x_{j-1}\over\dt}\right)^2
-4\kappa t_{3,j}{x_j-x_{j-1}\over\dt}\cdot s_j-2{\dt\over\lambda}t_{1,j}\\
&&-4\dt{e\over c\hbar}t_{3,j}s_j\cdot A_j\left({x_j+x_{j-1}\over2}\right)
+\dt{e\over c\hbar}V_j(x_j)+C\biggr],\nonumber
\eea
$x_0=x_i$, $\phi_0=\phi_i$, $\chi_0=\chi_i$, $x_N=x_f$,
$\phi_N=\phi_f$, $\chi_N=\chi_f$, 
$A_j(x)=A(x,j\tau/N)$, $V_j(x)=V(x,j\tau/N)$. We used the mid-point 
prescription in the vector potential to ensure gauge invariance and
the vectors corresponding to the spin and the chirality spaces
are given by
\bea
s_3&=s_{3,j}~~~~s_+&=\sqrt{1-s_{3,j}^2}e^{{i\over2}(\phi_j+\phi_{j-1}+2\pi n_j)}\nonu
t_3&=t_{3,j}~~~~t_+&=\sqrt{1-t_{3,j}^2}e^{{i\over2}(\chi_j+\chi_{j-1}+2\pi N_j)}.
\eea
By the shift of the variables
$\phi_j\to\phi_j-2\pi\sum_{k\le j}n_k$, 
$\chi_j\to\chi_j-2\pi\sum_{l\le j}N_k$, and the extension of the region
of integration for $\phi$ and $\chi$ over the whole real axis the summation over 
$n_j$ and $N_j$ decouples and can be ignored. 
The resulting path integral expression is
\be
G(x_f,\phi_f,\chi_f,x_i,\phi_i,\chi_i;\tau)=
\prod\limits_{k=1}^{N-1}\left(\int dx_kd\phi_kd\chi_k\right)
\prod\limits_{j=1}^N\left(\int{ds_{3,j}\over2\pi}{dt_{3,j}\over2\pi}\right)
e^{i\ca},
\ee
where
\bea
\ca&=&\sum\limits_{j=1}^N\biggl[\left(s_{3,j}+\hf\right)(\phi_j-\phi_{j-1})
+\left(t_{3,j}+\hf\right)(\chi_j-\chi_{j-1})\nonu
&&+{\kappa\over2}\left({x_j-x_{j-1}\over\dt}\right)^2
-4\kappa t_{3,j}{x_j-x_{j-1}\over\dt}\cdot s_j-2{\dt\over\lambda}t_{1,j}\\
&&-4\dt{e\over c\hbar}t_{3,j}s_j\cdot A_j\left({x_j+x_{j-1}\over2}\right)
+\dt{e\over c\hbar}V_j(x_j)
+C\biggr],\nonumber
\eea
and
\be
s_j=\pmatrix{\sqrt{1-s_{3,j}^2}\cos{\phi_j+\phi_{j-1}\over2}\cr
\sqrt{1-s_{3,j}^2}\sin{\phi_j+\phi_{j-1}\over2}\cr s_{3,j}},~~~
t_j=\pmatrix{\sqrt{1-t_{3,j}^2}\cos{\chi_j+\chi_{j-1}\over2}\cr
\sqrt{1-t_{3,j}^2}\sin{\chi_j+\chi_{j-1}\over2}\cr t_{3,j}}.\label{uvect}
\ee
The integration over the variable $\phi_k$, $\chi_k$ can be restricted
within the interval $[-2\pi,2\pi]$ due to the periodicity of the
integrand. One thus finds in the continuum limit the expression
\be
\int D[x]D[s]D[t]
e^{i\int d\tau\left\{{\kappa\over\dt}\left[\hf\left({dx\over d\tau}\right)^2
-4t_3{dx\over d\tau}\cdot s\right]-2{mc\over\hbar}t_1
-4{e\over c\hbar}t_3 s\cdot A(x)+{e\over c\hbar}V(x)+
\left(s_3+\hf\right){d\phi\over d\tau}
+\left(t_3+\hf\right){d\chi\over d\tau}\right\}}
\label{fpint}
\ee
for the solution of the Dirac equation, where the trajectories $s(\tau)$ and
$t(\tau)$ are on the sphere of unit length. 

Several remarks are in order at this point.
(a) As announced above, $D[s]=D[\cos\theta]D[\phi]$ and 
$D[t]=D[\cos\Theta]D[\chi]$ represent the integrations 
over the orientation of the spin and the chirality 
which are parametrized as
$s=(\sin\theta\cos\phi,\sin\theta\sin\phi,\cos\theta)$,
and $t=(\sin\Theta\cos\chi,\sin\Theta\sin\chi,\cos\Theta)$, respectively.
(b) Note that the half of the winding numbers $\hf\int d\phi$
and $\hf\int d\chi$ in the action generate
the sign $-1$ for rotations by $2\pi$. Thus the
path integration involves double valued amplitudes in the spin {\em and}
the chirality spaces. This is equivalent by saying that the integration 
$D[s]$ and $D[t]$ is over the covering space which contains $S_3$ twice,
in agreement with the mid-point prescription for $s$ and $t$, i.e.
the apparence of the half angles in \eq{uvect}.
(c) The variables $(\cos\theta,\phi)$ and $(\cos\Theta,\chi)$ can 
be considered as canonical momenta and coordinates.
The continuous formal notation actually hides the fact that there are
one less coordinate than momentum variable in the path integral.
The terms $s_3d\phi/d\tau$ and $t_3d\chi/d\tau$ 
are of the usual form $p\dot q$, characterizing the path integrals in
the phase space. The discrete spectrum of the angular momentum
and the corresponding finite integration range for $s_3$ and $t_3$
prevent us from eliminating the momentum variables and to reduce
the amplitude to a path integral in the coordinate space only. This
complication leaves behind a rather singular trajectory structure
for the angle variables.
(d) The term involving $t_1$ generates the rotation
of the chiral angle $\Theta$. In the massless case the $\chi$ integration
gives $dt_3/d\tau=0$, the conservation of the chiral angle
for massless fermions. The second and the fourth terms in the exponent
are responsible for the spin precession modulated by the
chiral angle. 
(e) Though the parameter $\kappa$ is left free we recall
that the saddle point trajectory $x_{cl}(\tau)$ of the path integral
\eq{pimatr} is the smoothest for $\kappa=\pi$ \cite{florence}. 
(f) It is worthwhile mentioning the double role the piece $O(\dot x^2)$
of the action plays: It generates a second order finite difference equation 
for the saddle point trajectories as expected from the semiclassical
limit and eliminates the species doubling by appearing as 
a Wilson term. Note the manifest chiral invariance.

Finally, a remark concerning the second quantization, mentioned at the 
introduction. Due to the spin-statistics theorem the phase
corresponding to the rotations by $2\pi$ and the exchange 
of equivalent particles are the same. Since the factor
$-1$ of the rotations is built in \eq{uvect} one hopes to arrive at the
second quantized path integral without Grassman variables.
Furthermore, the chiral invariance of the first quantized
amplitude \eq{fpint} suggests that the no-go theorem for
lattice chiral fermions \cite{nn} could be avoided.

\acknowledgments
I thank L. Schulman for helpful remarks in the course of this work.

\end{document}